\documentclass[doublecol]{epl2}
\usepackage{graphicx}
\usepackage{subfigure}
\usepackage{cite}
\usepackage{amssymb}
\usepackage{newlfont}
\usepackage{mathrsfs}
\usepackage{amsfonts}
\usepackage{stmaryrd}
\usepackage{booktabs}
\usepackage{tabularx}
\usepackage{algorithm}
\usepackage{algorithmic}

\title{Analysis of stability of community structure across multiple hierarchical levels}
\shorttitle{Analysis of stability of community structure across multiple hierarchical levels}

\author{Hui-Jia Li\inst{1} \and Xiang-Sun Zhang\thanks{Corresponding authors: \email{zxs@amt.ac.cn}}\inst{2,3}}

\shortauthor{Hui-Jia Li \etal}

\institute{
  \inst{1} School of Management Science and Engineering, Central University of Finance and Economics, Beijing 100080, China.\\
  \inst{2} Academy of Mathematic and Systems Science, Chinese Academy of Sciences, Beijing 100190, China.\\
  \inst{3} National Center for Mathematics and Interdisciplinary Sciences, Chinese Academy of Sciences, Beijing 100190, China.\\
}

\abstract{The analysis of stability of community structure is an important problem for scientists from many fields. Here, we propose a new
framework to reveal hidden properties of community structure by quantitatively analyzing the dynamics of Potts model.
Specifically we model the Potts procedure of community structure detection by a Markov process, which has a clear
mathematical explanation. Critical topological information regarding to multivariate spin configuration could also be inferred from the spectral
significance of the Markov process. We test our framework on some example networks and find it doesn't have resolute limitation problem at all. Results have shown the model we proposed is able to uncover hierarchical structure in different scales effectively and efficiently.}

\begin{document}

\maketitle

\section{1. Introduction}

Community structure detection \cite{Newman01, Newman02} is a main focus of
complex network studies. It has attracted a great deal of
attentions from various scientific fields. Intuitively, community refers
to a group of nodes in the network that are more densely connected
internally than with the rest of the network. A well known exploration for this problem is the concept of modularity, which
is proposed by Newman et al \cite{Newman01, Newman02} to quantify a network's partition. Optimizing modularity is effective for community
structure detection and has been widely used in many real networks.
However, as pointed out by Fortunato et al\cite{Fortunato},
modularity suffers from the resolution limit problem which is concerned
about the reliability of the communities detected through the
optimization of modularity. Complementary to the modularity concept, many efforts are devoted to understanding the
properties of the dynamical processes taking place in the underlying
networks. Specifically, researchers have begun to investigate the
correlation between the community structure and the
dynamical systems, such as synchronization\cite{Arenas}\cite{Arenas1} and random walk process\cite{Delvenne}\cite{Zhou}.

Potts dynamical model is a powerful tool which has been applied to
uncover the thermodynamical behaviors in
networks\cite{Blatt,Reichardt}. It models an inhomogeneous
ferromagnetic system where each node is viewed as a labeled spin in
the network. The configuration of the system is defined by the
interactions between the nodes.
Considering an unweighted network with $N$ nodes without self-loops, a
spin configuration $\{S\}$ is defined by assigning each node $i$ a
spin label $s_i$ which may take integer values $s_i=1,\cdots,K$. To
characterize the coherence between two nodes,
spin-spin correlation $C=(C_{ij})$ is defined as the thermal
average of $\delta_{s_i,s_j}$:
\begin{equation} \label{eq:1}
%\begin{split}
C_{ij}=\langle \delta_{s_i,s_j}\rangle,
%\end{split}
\end{equation}
which represents the probability that spin variables $s_i$ and $s_j$
have the same value. $C_{ij}$ takes values from the interval [0,1],
representing the continuum from no coupling to perfect accordance of
nodes $i$ and $j$. In section 4, we develop a novel hierarchical block model which can calculate $C$ efficiently.
The $C$ is corresponding to the average spin correlation across multiple level of the hierarchical structure.

\begin{figure}
\center
\includegraphics[width=6cm,height=3.8cm,angle=0]{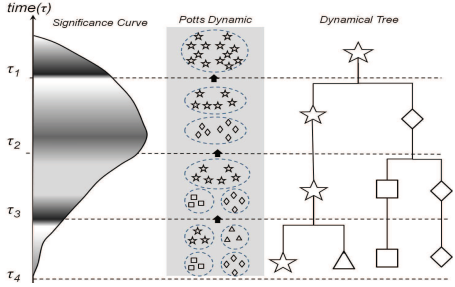} \caption{
Dynamics of spin configuration of four communities when they go
through several local uniform states to the global stable state.
Different spin values are described by different shapes in dynamical
tree. $\tau_i$ represents the time at which the system has $i$ different spin states}. The significance curve of spin configurations at different
times can be calculated shown in the left part of the graph.
\label{fig.1}
\end{figure}

If the system is not homogeneous but has a community structure, the
states are not just ferromagnetic or paramagnetic\cite{Blatt}. We assume that
the spins will go through a hierarchy of local uniform states
(meta-stable states) as time increases which is shown in Fig.\ref{fig.1}, before they
reach a globally stable state with the same value. In each local uniform state, spin values of nodes within
the same communities are identical. Correspondingly, one can
calculate the hitting and exiting time of each local uniform state
and there should be a big gap between them when a well-formed community structure exists. The significance of spin configurations at different time can also be calculated to illustrate the amplitude of variation.

In this letter, using the Potts model and spectral theory, we firstly uncover the relationship between
community structure of a network and its meta-stability of spin
dynamics, and then propose the significance of communities to
characterize and analyze the underlying spin configuration. For any
given network, one can straightforwardly get critical information
related to its community structure, such as the stability and the optimal number of communities across multiple
timescales without using particular partition algorithms. We then use
phase transition of stochastic dynamical system to prove that the stability we proposed is able to
indicate the significance of community structure more theoretically which is based on eigengap theory. Furthermore, a novel hierarchical
block model is proposed which can calculate spin correlation at each layer of the network structure. Finally, we test our framework on some examples of complex networks. Results show the model we proposed is able to uncover the hierarchical
structure in different scales effectively and efficiently and doesn't have resolute limitation problem at all.

\section{2. The framework}

In order to establish the connection between the community structure and the
local uniform behavior of Potts model, we introduce a Markov
stochastic model featured by spectral significance for the
network. Markov process is a useful tool and has been applied to
find communities\cite{Delvenne}. Let $P=(p_{ij})$ be the stochastic transition matrix and the element $p_{ij}$ is defined as

\begin{equation} \label{eq:2}
%\begin{split}
p_{ij}=\frac{C_{ij}}{\sum_{j=1}^NC_{ij}}
%\end{split}
\end{equation}
where $C_{ij}$ is the spin-spin correlation function defined in Eq.(\ref{eq:1}). Via this representation, the tools of stochastic
theory and finite-state Markov processes
\cite{Delvenne} can be utilized for the purpose of
community structure analysis.

For this ergodic Markov process, $P^t$ represents the transition probability matrix between nodes over a period of $t$ time
steps. To compute the transition matrix $P^t$, the eigenvalue
decomposition of $P$ is used. If $\lambda_k$ with $k =1,\cdots,n$
denote the eigenvalues of $P$, and its right and left eigenvectors
$u_k$ and $f_k$ are scaled to satisfy

\begin{equation}\label{eq:3}
Pu_{k}=\lambda_{k}u_{k}, f_{k}P=\lambda_{k}f_{k}
\end{equation}.

The orthonormality relation of $u_k$ and $f_l$ is satisfied:

\begin{equation} \label{eq:4}
%\begin{split}
u_kf_l=\delta_{kl},
%\end{split}
\end{equation}
and the spectral representation of $P$ is given by

\begin{equation} \label{eq:5}
P^{t}=\sum_k\lambda^{t}_ku_kf_k
\end{equation}
We assume that eigenvalues of $P$ are sorted such that $\lambda_1 =1
>|\lambda_2|\geq|\lambda_3|\geq ... \geq|\lambda_{n}|$. Because of the largest eigenvalue $\lambda_1=1$, when time
$t\rightarrow \infty$,
$P^{(0)}=P^{\infty}=u_1f_1$. The
convergence of every initial distribution to the stationary
distribution $P^{(0)}$ corresponds to the fact that the spins of
whole system ultimately reach exactly the same value, as
time increases. This perspective belongs to
a timescale $t\rightarrow\infty$, at which all eigenvalues
$\lambda^{t}_k$ go to 0 except for the largest one,
$\lambda^{t}_1=1$. In the other extreme of a timescale $t =0$, $P^t$
becomes the stationary distribution matrix. All of its columns are
different, and the system disintegrates into as many spin values as possible. Then, we simply extend $P^t$ to the symmetrical
form $G^{(t)}=(P^t+(P^t)^{T})/2$.

Suppose the partition method divides the network $A$ into $K$
communities or sets $V_k\subset V, k\in\{1,2,\cdots, K\}$ which are disjoint
and the sets $V_1$, $V_2$,..., $V_K$ together form a partition of
node set $V$. The number of nodes in each community is denoted by $N_k=|V_k|$. We
take the time series into consideration. Therefore, we define
the significance of a given community $k$ by the ratio of inner
correlations as

\begin{equation} \label{eq:6}
%\begin{split}
S_k^{(t)}=\sum_{i,j\in V_k}\frac{[G^{(t)}]_{i,j}}{N_k}
%\end{split}
\end{equation}
$S_k^{(t)}$ can be viewed as a function of timescale $t$ and we can
use it to study the trend of community structure as time goes on.

Further discussion is facilitated by reformulating the average
association objective in matrix form. We denote the membership
vector of community $k$ by $x_k$, a binary vector that describes
each node's involvement in community $k$. The hard
partition and disjointness of sets $V_k$ requires that the
vectors $x_l$ and $x_s$ are orthogonal. Given the number of communities $K$,
the communities are found by maximizing the objective function

\begin{equation} \label{eq:7}
%\begin{split}
J_K^{(t)}=\sum_{k=1}^K\sum_{i,j\in V_k}\frac{[G^{(t)}]_{i,j}}{N_k}=\sum_{k=1}^K\frac{x_k^{T} G^{(t)}x_k}{x_k^T x_k}
%\end{split}
\end{equation}

The objective is to be maximized under the conditions $x_k\in
\{0,1\}$ and $x_l^T x_s=0$ if $l\neq s$. Eq.(\ref{eq:7}) can be
rewritten as a matrix trace by accumulating the vectors $u_k$ into a
matrix $X=(x_1, x_2,...,x_K)$. We can then write the objective
$J_K^{(t)}$ as

\begin{equation} \label{eq:8}
%\begin{aligned}
\begin{array}{lcl}
J_K^{(t)}=tr\{(X^T X)^{-1}X^T G^{(t)}X\} \\
=tr\{(X^T X)^{-1/2}X^T G^{(t)}X(X^T X)^{-1/2}\}
\end{array}
%\end{aligned}
\end{equation}
where matrix $X^T X$ is diagonal. The substitution $Y=X(X^T
X)^{-1/2}$ simplifies the optimization problem to $J_K^{(t)}=tr\{
Y^TG^{(t)}Y\}$. The condition $Y^TY=I_K$ is automatically satisfied
since

\begin{equation} \label{eq:9}
%\begin{split}
Y^T Y=(X^T X)^{-1/2}(X^T X)(X^T X)^{-1/2}=I_K.
%\end{split}
\end{equation}

The vectors $y_k$ thus have unit length and are orthogonal to each
other. The optimization problem can be written in terms of the
matrix $Y$ as

\begin{equation} \label{eq:10}
%\begin{split}
\max_{Y^T Y=I}tr\{Y^T G^{(t)}Y\}.
%\end{split}
\end{equation}

According to Rayleigh-Ritz theorem\cite{Shi}, the maximum for
this problem is attained when columns of $Y$ is the right eigenvectors $U=\{u_1,...,u_K\}$
corresponding to the $K$ largest eigenvalues of the symmetric
correlation matrix $G^{(t)}$. Then the strength of such a
community is approximately equal to its corresponding $t$-th power of the eigenvalue

\begin{equation} \label{eq:11}
%\begin{split}
S_k^{(t)}\approx\frac{u_k^T G^{(t)}u_k}{u_k^T u_k}=\lambda_k^t\frac{u_k^T
u_k}{u_k^T u_k}=\lambda_k^t
%\end{split}
\end{equation}

For the convergence of the Potts model across multiple timescales,
the vanishing of the smaller eigenvalues as the time growing
describes the loss of different spin states and the removal of the
structural features encoded in the corresponding weaker
eigenvectors. For the purpose of community identification,
intermediate timescales of local uniform states are interesting. If
we want to identify $z$ communities, we expect to find $P^t$ at a
given timescale, the eigenvalues $\lambda^t_k$ may be significantly
different from zero only for the range $k=1,...,z$. This is achieved
by determining $t$ such that $|\lambda_k|^t\thickapprox0$.

From another perspective, because the eigenvalues are sorted by
$\lambda_1 =1 >|\lambda_2|\geq|\lambda_3|\geq ...
\geq|\lambda_{n}|$, the strength of a community at time $t$,
$\lambda_k^t$, can also be viewed as the robustness of $k$-spin
state at time $t$. At this point, the eigengap
$\lambda_{k-1}^t-\lambda_k^t$ can be interpreted as the
``difficulty'' that the $k$-spin state transfer to the $(k-1)$-spin
state at time $t$. The number of
communities $\Lambda$ at time $t$ is then inferred from the location
of the maximal eigengap, and this maximal value can be used as a
quality measure for the most stable state. The $\Lambda (t)$ is
formally defined as

\begin{equation} \label{eq:12}
%\begin{split}
\Lambda(t)=arg[max_k(\lambda_{k-1}^t-\lambda_k^t)]
%\end{split}
\end{equation}
From a global perspective if the number of communities $\Lambda$ doesn't
change for the longest time, we can consider it as the optimal
number for this network, represented as $\Psi$.

%%虽然最优社团数目持续不变，但是内部的变化却不是透明的。
To a certain
extent, the most stable state can represent the spin configuration
of the whole network. Thus, we define the stability of community
structure at each timescale, $\Theta(t)$, as the stability of the
most stable spin state:

\begin{equation} \label{eq:13}
%\begin{split}
\Theta(t)=\lambda_{\Lambda(t)-1}^t-\lambda_{\Lambda(t)}^t
%\end{split}
\end{equation}
Our expectation is that from the trend of $\Theta(t)$, one can
find the most stable timescale for community structure where
$\Theta(t)$ reaches the maximal. Furthermore, from a global perspective, we can use the largest
stability corresponding to $q$ communities,
$\Gamma(q)=max\{\Theta(t)|\Lambda(t)=q\}$, to indicate the
robustness of a network, defined as the stability of the structure
with $q$ communities. While $\Gamma(q)$ tries to directly
characterize the network structure rather than a specific network
partition thus very convenient to estimate the modularity
property of the network.

\section{3. Prove the validity of stability}

Many measures have been defined to indicate the significance of community structure, such as modularity $Q$ proposed by Newman et al\cite{Newman01}\cite{X}and spectral cut metrics\cite{Shi}. In\cite{Capocci}, the eigenvector of transition matrix $P$ is also found able to indicate the partition of the nodes in the network. The components of eigenvector corresponding to nodes within the same community have very similar values and the eigenvalue gaps between different communities can represent the significance of the modularity structure. In this part, we use phase transition of stochastic dynamic system to prove that the stability we proposed is in proportion to the eigenvalue gap. Thus, the larger the stability, the larger the significance of the community structure.

Let us demonstrate our argument for the simplest case that a network owning 2 communities.
As $\lambda_{1}\equiv 1$, according to Eq.(\ref{eq:5}), we write
\begin{equation}\label{eq:14}
P^{t}=u_{1} f_{1}+\lambda_{2}^{t} u_{2} f_{2}+\mid\lambda_{3}\mid^{t} B^{t},
\end{equation}
where $B^{t}$ is the remainder matrix
\begin{equation}\label{eq:15}
B^{t}=\sum_{k\geq 3}{\frac{\lambda_{k}^{t}}{\mid\lambda_{3}\mid^{t}}u_{k}f_{k}}
\end{equation}

Although $P$ need not to be diagonalizable,  the representation is guaranteed by the nondegeneracy of $\lambda_{1}$ and $\lambda_{2}$. Because $\sum_{x}f_{2}(x)u_{1}(x)=0$ and $u_{1}>0$, we deduce that
\begin{equation}\label{eq:16}
f_{M}\equiv max_{x\in X}f_{2}(x)>0>f_{m}\equiv min_{x\in X}f_{2}(x)
\end{equation}
Let us now fix $a>0$ and according to Eq.(\ref{eq:16}), we consider the two nodes sets belonging to two different communities:
\begin{equation}\label{eq:17}
\begin{array}{lcl}
I_{M}(a)=\{x\in X \mid\frac{f_{M}-f_{2}(x)}{f_{M}}<a\},\\
I_{m}(a)=\{x\in X \mid\frac{f_{m}-f_{2}(x)}{f_{m}}<a\},
\end{array}
\end{equation}
These two sets will in fact turn out to be the two phases. We also define $E_{M}(a)\equiv\varphi I_{M}(a)$, $E_{m}(a)\equiv\varphi I_{m}(a)$(where $\varphi$ indicates the complement of a set). We take $a<1$ and note that
\begin{equation}\label{eq:18}
%\begin{aligned}
a<1~\Rightarrow  I_{M}(a)~and~I_{m}(a)~are~disjointed.
%\end{aligned}
\end{equation}

Accordingly to the two nodes sets, the corresponding the two phases of the system are,
\begin{equation}\label{eq:19}
\begin{array}{lcl}
u_{M}^{t}(x)=u_{1}(x)+\lambda_{2}^{t}u_{1}(x)f_{M}+\mid\lambda_{3}^{t}\mid B_{xM}^{t},\\
u_{m}^{t}(x)=u_{1}(x)+\lambda_{2}^{t}u_{1}(x)f_{m}+\mid\lambda_{3}^{t}\mid B_{xm}^{t},
\end{array}
\end{equation}
Here $B_{xM}^{t}$ (respectively $B_{xm}^{t}$) is the value of $B_{xy}^{t}$ for some point $y(\equiv y_{M})$ (respectively, $y_{m}$) such that $f_{2}(y)=f_{M}$ (respectively, $f_{m}$). From Eq.(\ref{eq:19}), we have
\begin{equation}\label{eq:20}
\begin{array}{lcl}
u_{1}(x)=\frac{f_{M}u_{m}^{t}(x)-f_{m}u_{M}^{t}(x)}{f_{M}-f_{m}}+\mid\lambda_{3}^{t}\mid\frac{f_{m}B_{xM}^{t}-f_{M}B_{xm}^{t}}{f_{M}-f_{m}}\\
u_{2}(x)=\frac{f_{M}^{t}(x)-u_{m}^{t}(x)}{\lambda_{2}^{t}(f_{M}-f_{m})}+\frac{\mid\lambda_{3}^{t}\mid }{\lambda_{2}^{t}}\frac{B_{xm}^{t}-B_{xM}^{t}}{f_{M}-f_{m}}
\end{array}
\end{equation}
More generally, for any $y$ one can define
\begin{equation}\label{eq:21}
u_{y}^{t}(x)=u_{1}(x)+\lambda_{2}^{t}u_{2}(x)f_{2}(y)+\mid\lambda_{3}^{t}\mid B_{xy}^{t}\equiv P_{xy}^{t}
\end{equation}

We take the scalar product of $u_{M}^{t}$ and $u_{m}^{t}$ with $f_{2}$. There are
\begin{equation}\label{eq:22}
\sum_{x}f_{1}(x)B_{xy}^{t}=0.
\end{equation}
This follows from
\begin{equation}\label{eq:23}
B^{t}=\sum_{k\geq 3}\frac{\lambda_{k}^{t}}{\mid\lambda_{3}^{t}\mid} u_{k}f_{k}
\end{equation}
and therefore
\begin{equation}\label{eq:24}
\sum_{x}f_{2}(x)u_{M}^{t}(x)=\lambda_{2}^{t}f_{M},
\sum_{x}f_{2}(x)u_{m}^{t}(x)=\lambda_{2}^{t}f_{m},
\end{equation}

From Eq.(\ref{eq:24}) and the fact that $\sum{u_{M}^{t}(x)}=\sum{u_{m}^{t}(x)}=1$ we can deduce the basic identity
\begin{equation}\label{eq:25}
\lambda_{1}^{t}-\lambda_{2}^{t}=\sum_{x}u_{M}^{t}(x)(1-\frac{f_{2}(x)}{f_{M}})=\sum_{x}u_{M}^{t}(x)(1-\frac{f_{2}(x)}{f_{m}})
\end{equation}
The foregoing identity is a fundamental statement about the structure of the two communities. From the definition of $I_{M}(a),E_{M}(a)$ and the identify, we deduce
\begin{equation}\label{eq:26}
\sum_{x\in I_{M}(a)}u_{M}^{t}(x)>1-\frac{1-\lambda_{2}^{t}}{a},
\sum_{x\in E_{M}(a)}u_{M}^{t}(x)<\frac{1-\lambda_{2}^{t}}{a}
\end{equation}
with analogous inequalities for $M$ replaced by $m$

We have seen that
\begin{equation}\label{eq:27}
u_{2}(x)=\frac{u_{M}^{t}(x)-u_{m}^{t}(x)}{\lambda_{1}^{t}(f_{M}-f_{m})}+r(x),
\end{equation}
where $r(x)=\lambda_{2}^{t}$ $\times$(linear combination of $B_{xM}^{t}$,$B_{xm}^{t}$). From the basic identity in Eq.(\ref{eq:25}), we deduce
\begin{equation}\label{eq:28}
\begin{array}{lcl}
\sum_{x\in I_{M}}u_{2}(x)\geq\frac{1}{a[1-(\lambda_{1}^{t}-\lambda_{2}^{t})](f_{M}-f_{m})}+\epsilon,\\
\sum_{x\in I_{m}}u_{2}(x)\geq\frac{-1}{a[1-(\lambda_{1}^{t}-\lambda_{2}^{t})](f_{M}-f_{m})}+\epsilon^{,},
\end{array}
\end{equation}
where $\epsilon,\epsilon^{,}$ are $o(1-\lambda_{2}^{t})$.

From Eq.(\ref{eq:28}), we notice that the gap of eigenvector value is $\sum_{x\in I_{M}}u_{2}(x)-\sum_{x\in I_{m}}u_{2}(x)\geq\frac{2}{a[1-(\lambda_{1}^{t}-\lambda_{2}^{t})](f_{M}-f_{m})}+o(1-\lambda_{2}^{t})$ and its lower bound increases with the stability $\Theta(t)=\lambda_{1}^{t}-\lambda_{2}^{t}$. Thus, we declare that larger stability of communities will extend the eigenvector gap between them and thus enhance the significance of the community structure. Furthermore, one can easily extend the stochastic dynamic system to $k$-state, $k\geq 2$, in which the stability $\Theta(t)=\lambda_{k}-\lambda_{k+1}$ can also indicate the significance of structure owning $k$ communities.

\section{4. Estimate the spin correlation}

The spin correlation matrix $C$ is very important for the Potts dynamic. In this section, we propose a novel way to calculate $C$ using a new hierarchical block model method based on different granularity(resolution). Stochastic block model \cite{Karrer} is a useful tool to detect communities from networks or dynamical networks. However, the existing block model methods are restricted to the specific task of community detection and not suitable to models which need to extract multiple levels structure of hierarchical networks. In this part, the stochastic block model is extended to a multilevel form and exactly coincides with the dynamical process of the Potts model.

Let $A_{n\times n}$ be the adjacent matrix of network $N$, where $n$ is the number of nodes. Suppose all nodes of $N$ are divided into $L(1\leq L\leq n)$ blocks, denoted by $B_{n\times L}$, where $b_{il}=1$ if node $i$ is in block $l$, otherwise $b_{il}=0$. When each block is considered to be inseparable, the granularity of network $N$ can be measured by the number of blocks $g=L$. As $g$ decreases from $n$ to $1$, the granularity of $N$ degenerates from the finest to the coarsest. Let $B_{g}$ denotes the block matrix $B$ with a granularity $g$. In particular, we have $B_{1}=I_{n\times n}$. Let matrix $Z_{L\times K}(1\leq K\leq L)$ denotes such communities, where $K$ is the community number and $z_{lk}=1$ if block $l$ is labeled by community $k$, otherwise $z_{lk}=0$. Given $Z$, define $\Xi_{K\times n}$, where $\xi_{kj}$ denotes the probability that any node out of community $k$ expects to couple with node $j$; and define $\Omega(\omega_{1},...,\omega_{K})^{T}$, where $\omega_{k}$ denotes the prior probability that a randomly selected node will belong to community $k$. It is easy to show that spin correlation matrix at level $g$ is $C_g=B_{g}Z\Xi$ so calculating $C_g$ is corresponding to estimate $B_g$, $Z$ and $\Xi$.

Define $X=(K,Z,\Xi,\Omega)$ be a pattern unit of network $N$ with respect to $B_{g}$. According to the principle of maximizing the posterior
probability, the optimal $X$ for a given network $N$ under $B_{g}$ will be one with the maximal posterior probability. Moreover, we have
\begin{equation} \label{eq:29}
P(X|N,B_{g})\propto P(N|X,B_{g})P(X|B_{g})
\end{equation}
where $P(X|N,B_{g})$, $P(N|X,B_{g})$, and $P(X|B_{g})$ denote the posteriori of $X$ given $N$ and $B_{g}$, the likelihood of $N$ given $X$ and $B_{g}$, and the priori of $X$ given $B_{g}$, respectively.

As discussed above, an optimal $X$ will be the one with the maximal $P(X|N,B_{g})$ and to maximize $P(X|N,B_{g})$ is to maximize the product of $P(N|X,B_{g})$ and $P(X|B_{g})$. For a given $K$, the term $P(X|B_{g})$ is a constant, and thus, to maximize $P(X|N,B_{g})$ is to maximize $L(N|X,B_{g})$.

Let $L(N|X,B_{g})=lnP(N|X,B_{g})$, and we have
\begin{equation} \label{eq:30}
L(N|X,B_{g})=\sum_{l=1}^{L}{\sum_{b_{il}\neq 0}{\sum_{k=j}^{K}{\prod_{j=1}^{n}{f(\xi_{kj},a_{ij})w_{k}}}}}
\end{equation}
where $f(x,y)=x^{y}(1-x)^{1-y}$.

Considering the expectation of $L(N,Z|X,B_{g})$ on $Z$, we have:
\begin{equation} \label{eq:31}
\begin{array}{lcl}
E[L(N,Z|X,B_{g})]\\
=\sum_{l=1}^{L}\sum_{b_{il}\neq 0}\sum_{k=1}^{K}\gamma_{lk}(\sum_{j=1}^{n}(\ln f(\xi_{kj},a_{ij}))+\ln \omega_{k})
\end{array}
\end{equation}
where $E[z_{lk}]=\gamma_{lk}=P(y=k|b=l,X,B_{g})$, i.e., the probability of block $l$ will be labeled as community $k$ given $X$ and $B_{g}$.
Let $J=E[L(N,Z|X,B_{g})]+\lambda(\sum_{k=1}^{K}w_{k}=1)$, we have:
\begin{equation} \label{eq:32}
\left \{ \begin{array}{l}
\frac{\partial{J}}{\partial{\xi_{kj}}}=0\\
\frac{\partial{J}}{\partial{\omega_{k}}}=0\\
\frac{\partial{J}}{\partial{\lambda}}=0\\
\end{array}
\right. \Rightarrow\left \{ \begin{array}{l}
\xi_{kj}=\frac{\sum_{l=1}^{L}{\sum_{b_{il}\neq 0}{a_{ij}\gamma_{lk}}}}{\sum_{l=1}^{L}{\sum_{b_{il}\neq 0}{\gamma_{lk}}}} \\
\omega_{k}=\frac{\sum_{l=1}^{L}{\sum_{b_{il}\neq 0}{\gamma_{lk}}}}{\sum_{k=1}^{K}\sum_{l=1}^{L}{\sum_{b_{il}\neq 0}{\gamma_{lk}}}}\\
=\frac{\sum_{l=1}^{L}{\sum_{b_{il}\neq 0}{\gamma_{lk}}}}{n}
\end{array}\right.
\end{equation}

Let $P(y=k|v=i)$ be the probability that node $i$ belongs to community $k$ given $X$ and $B_{g}$, We have:
$\gamma_{lk}=P(y=k|b=l,X,B_{g})=\sum_{b_{il}\neq0}\frac{1}{\sum_{i=1}^{n}b_{il}}P(y=k|v=i)$
where $\frac{1}{\sum_{i=1}^{n}b_{il}}$ is the probability of selecting node $i$ from block $l$.
According to the Bayesian theorem, we have:
\begin{equation} \label{eq:33}
P(y=k|v=i)=\frac{P(y=k)P(v=i|y=k)}{\sum_{k=1}^{K}P(y=k)P(v=i|y=k)}.
\end{equation}
and
\begin{equation} \label{eq:34}
P(y=k)P(v=i|y=k)=\prod_{j=1}^{n}{f(\xi_{kj},a_{ij})\omega_{k}}
\end{equation}

Thus
\begin{equation} \label{eq:35}
%\begin{split}
\gamma_{lk}=\frac{1}{\sum_{i=1}^{n}{b_{il}}}\times \sum_{b_{il}\neq 0}{\frac{\prod_{j=1}^{n}{f(\xi_{kj},a_{ij})w_{k}}}{\sum_{k=1}^{K}{\prod_{j=1}^{n}{f(\xi_{kj},a_{ij})w_{k}}}}}
%\end{split}
\end{equation}
As a conclusion, a local optimum of maximizing Eq.(\ref{eq:29}) will be guaranteed by recursively calculating Eq.(\ref{eq:32})and Eq.(\ref{eq:35}) with granularity $g$. An optimal pattern unit $X=(K,Z,\Xi,\Omega)$ is calculated given $B_g$ and consequently spin correlation matrix $C_g=B_gZ\Xi$.

For a given network, the hierarchical calculation process of $C_g$ as $g$ decreases from $n$ to 1 can be incrementally proceeded as follows:
First, constructing the ground layer by taking each node as one block, and $L=g=n, B_g=I_{n\times n}$. Thus, $C_n=B_gZ\Xi=A$. Then clustering it into $n-1$ communities by selecting a model $X_{n}$ with a maximum $P(X_{n}|N,B_{n})$.
Second, according to $X_{n}$, form $B_{n-1}$ by capsuling each cluster in the ground layer as one block. $B_{n-1}=B_n\times Z_{n-1}$ and $C_{n-1}=B_{n-1}Z\Xi=A$. Then clustering these $n-1$ blocks into $n-2$ communities by calculating a new model $X_{n-1}$ with a maximum $P(X_{n-1}|N,B_{n-1})$. Repeat the second step to construct more layers until the process converges, i.e., all blocks are grouped into only 1 communities. After calculating all $n$ layers $C_g$, the average spin correlation matrix $C$ can be taken as the hierarchical average $C=\langle C_g\rangle$, $g=1,..,n$. One can easily find that the process coincides with the hierarchical dynamical process of Potts model described in Fig.\ref{fig.1}.

\section{5. Experiments}

\begin{figure}
\center
  \subfigure[]{
    \label{fig:subfig:2a} %% label for second subfigure
    \setcounter{subfigure}{1}
    \includegraphics[width=4cm,height=2.8cm]{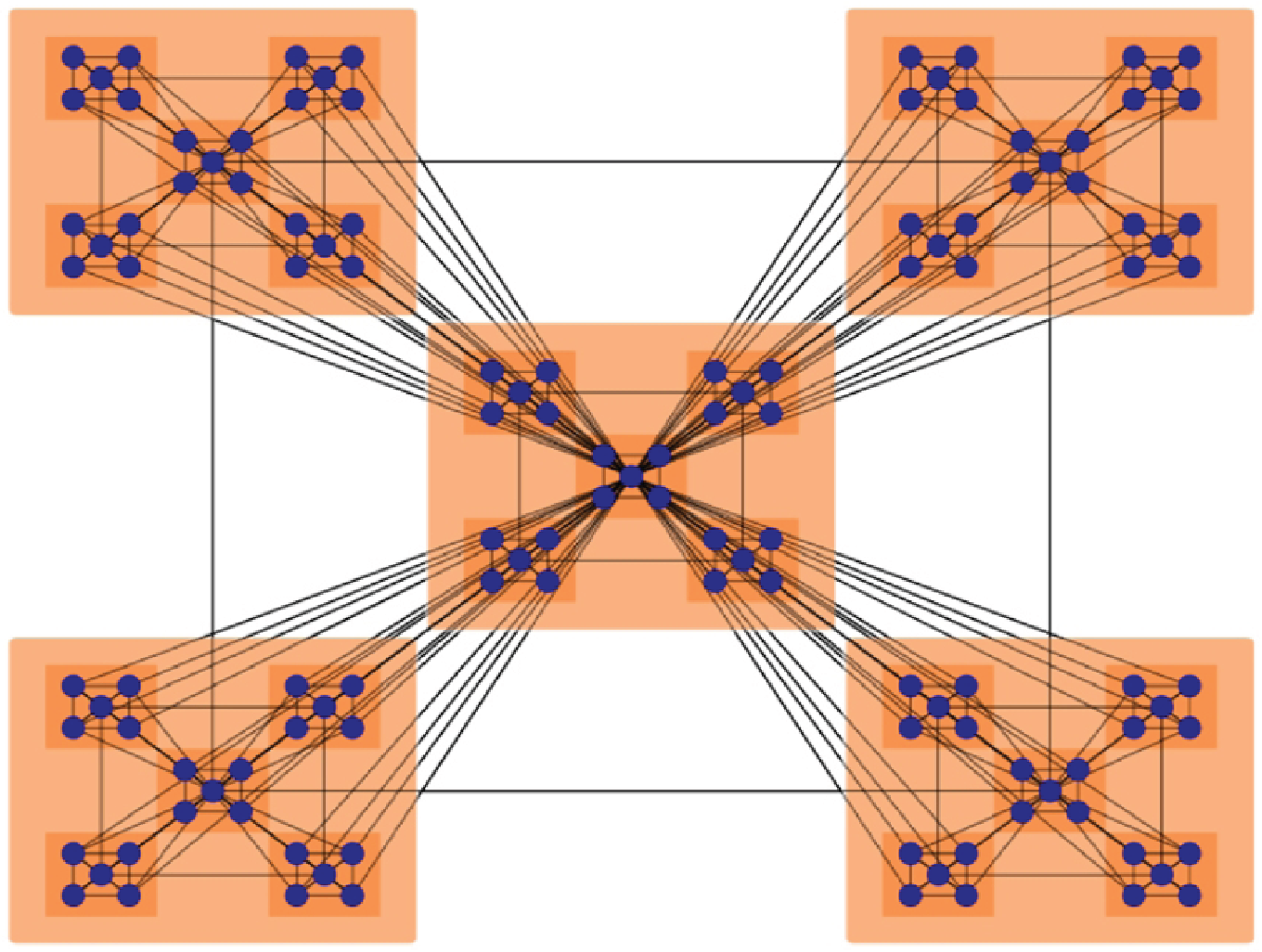}}
  \subfigure[]{
    \label{fig:subfig:2b} %% label for second subfigure
    \setcounter{subfigure}{2}
    \includegraphics[width=6cm,height=3.4cm]{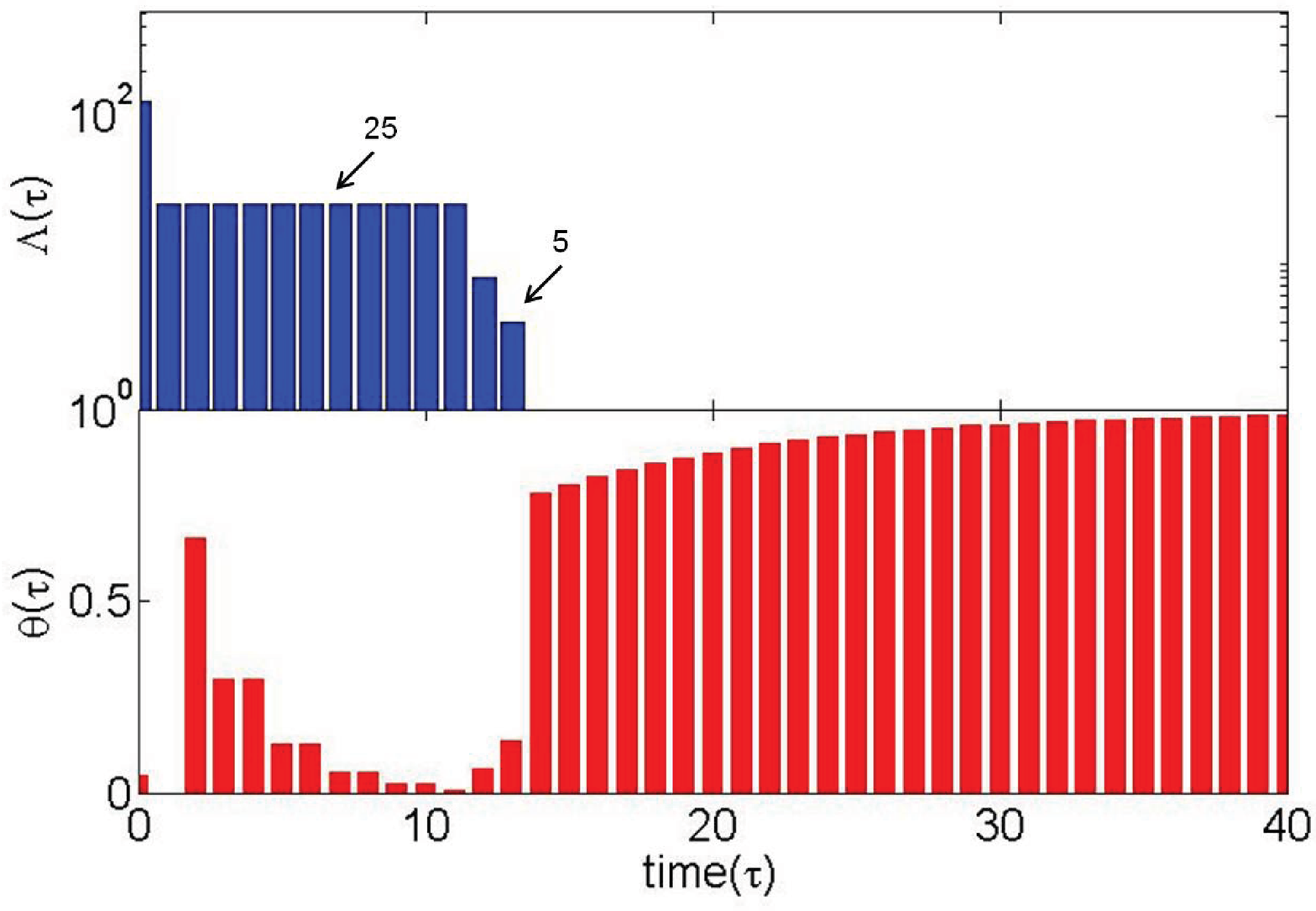}}
\caption{(a) Structure of $RB125$, with 25 dense
communities and 5 sparse communities, are highlighted in the original
network. (b) The value of $\Lambda(\tau)$ and $\Theta(\tau)$ versus
time $\tau$.} \label{fig.2}
\end{figure}

\begin{figure}
\center
  \subfigure[]{
    \label{fig:subfig:3a} %% label for second subfigure
    \setcounter{subfigure}{1}
    \includegraphics[width=4cm,height=2.8cm]{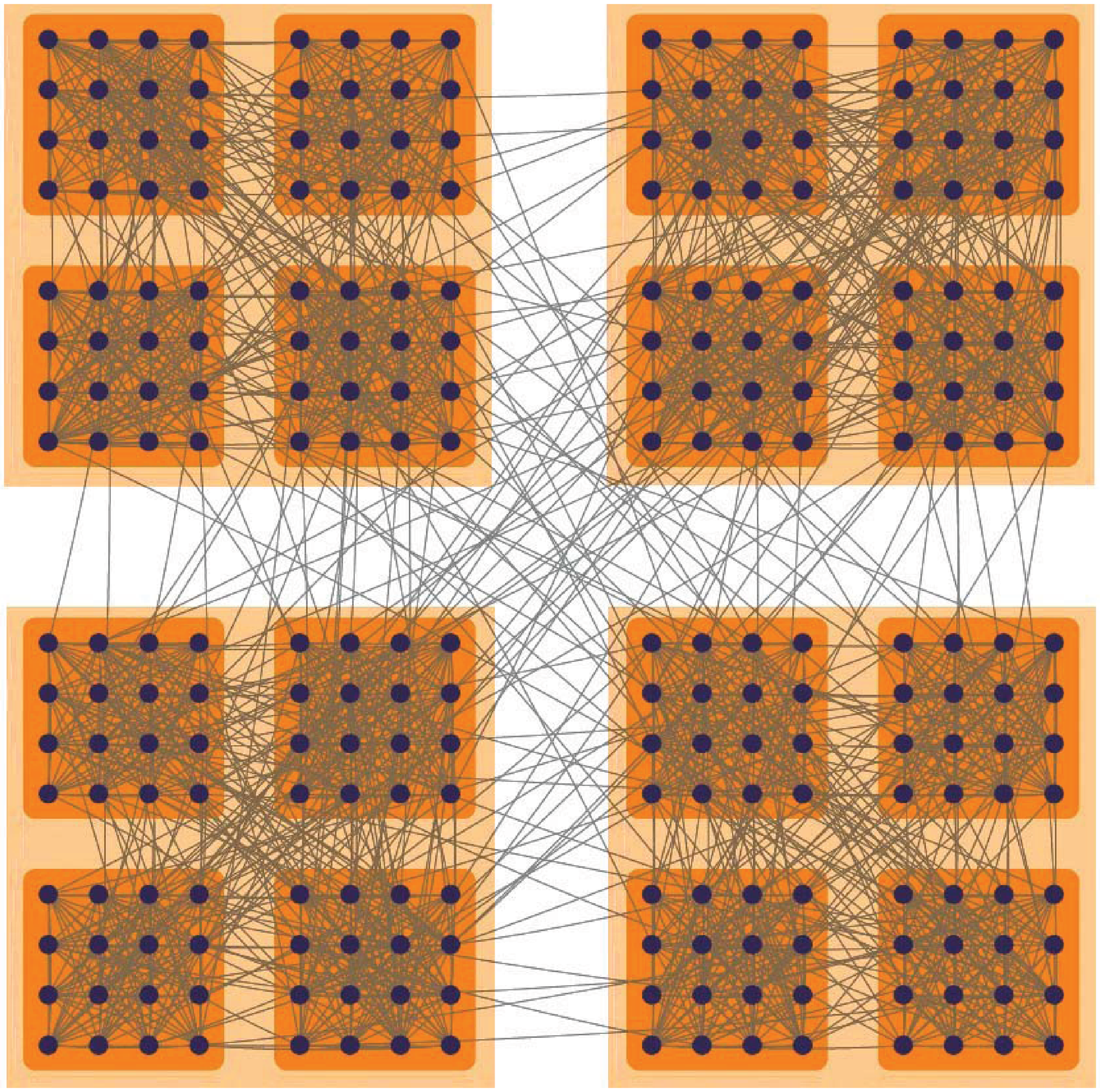}}
  \subfigure[]{
    \label{fig:subfig:3b} %% label for second subfigure
    \setcounter{subfigure}{2}
    \includegraphics[width=6cm,height=3.4cm]{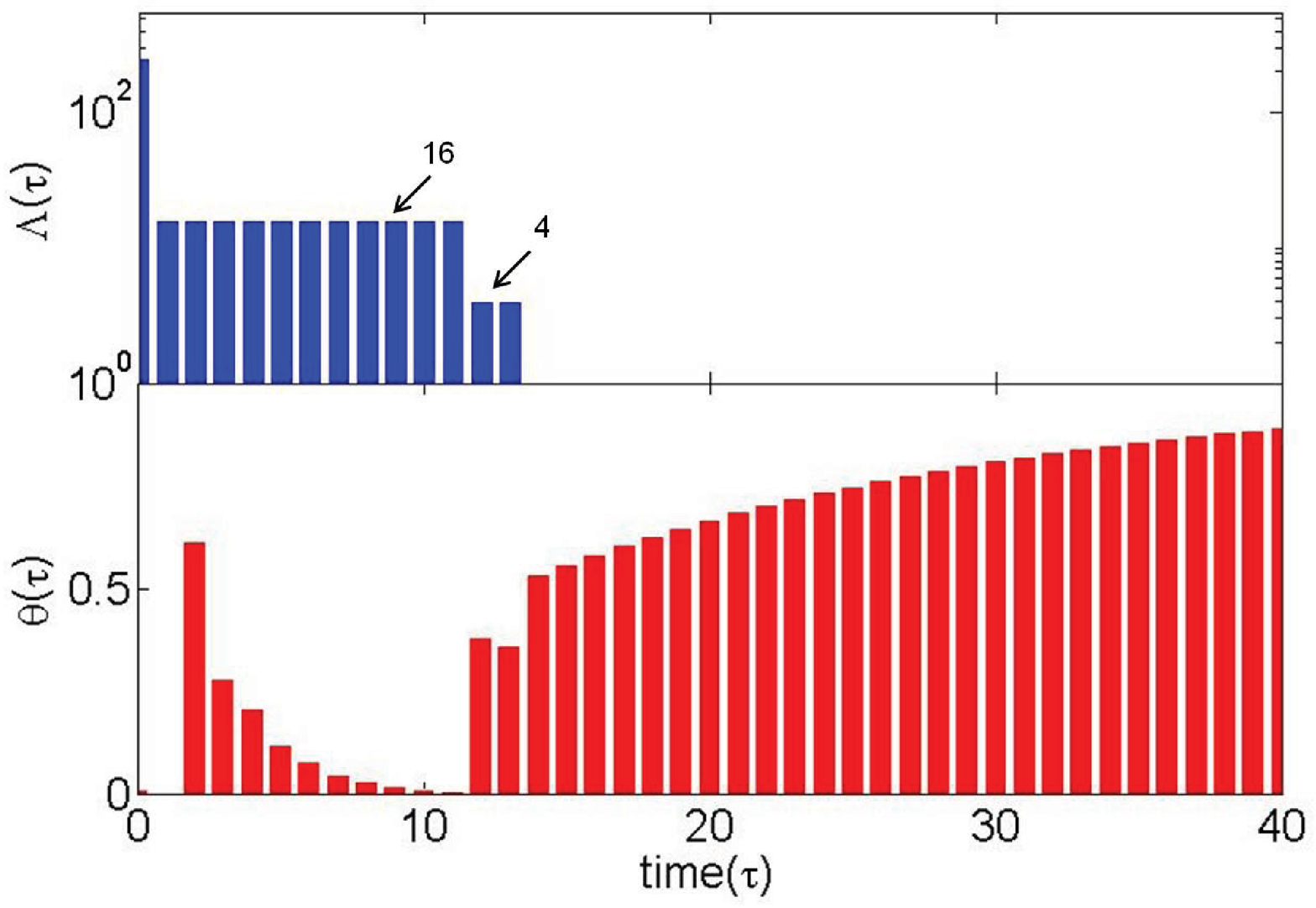}}
\caption{(a) Structure of $H13$-$4$, with 16 dense
communities and 4 sparse communities, are highlighted in the original
network. (b) The value of $\Lambda(\tau)$ and $\Theta(\tau)$ versus time $\tau$.}
\label{fig.3}
\end{figure}

To show that the model can uncover hierarchical structures in
different scales, Fig.\ref{fig.2} and Fig.\ref{fig.3} give two
examples of the multi-level community structures, $RB125$ network\cite{Ravasz} and $H13$-$4$ network\cite{Arenas}.
In both examples, the most persistent $\Lambda$ reveals the actual
number of hierarchical levels hidden in a network. The significance of
such levels can be quantified by their corresponding length of
persistent time. Longer the time persists, more robust the
configuration is. From Fig.\ref{fig:subfig:2b} and
Fig.\ref{fig:subfig:3b}, we can observe 25 and 16 are the optimal
numbers of communities in $RB125$ and $H13$-$4$ networks owning the
longest persistence, respectively. However, 5 modules and 4 modules
are also reasonable partitions which show another fuzzy level of the
hierarchical networks. These results are perfectly consistent
with the generating mechanisms and hierarchical patterns of these two
networks.

Furthermore, we also show that the variation tendency of stability
$\Theta(\tau)$ in the two cases shed a light on the spin
configuration. From Fig.\ref{fig:subfig:2b} and
Fig.\ref{fig:subfig:3b}, there are
some local maximal values representing better community
structure. Thus, we can find these local maximal timescales $^\tau$
corresponding to the desirous number of communities and apply $G^\tau$
to a specific partition method. Furthermore, the stability will reach the lowest value at the end time of all
$\Lambda$. The stability begins to increase when it transits to a new state. One can
use $\Theta(\tau)$ to estimate the modularity property of complex
networks, and the larger the $\Theta$ is, the stronger the network
community structure. So, one can find the largest corresponding $\Theta$ value for a specific number of community
$\Lambda$ and use it to indicate the robustness of modularity structure. For
$H13$-$4$ shown in Fig.\ref{fig:subfig:3b}, the stability of 16
communities structure, $\Gamma(16)=0.62$ when $\tau=3$, is larger than
$\Gamma(4)=0.43$ when $\tau=12$. This indicates that the community
structure containing 16 modules is more robust than community
structure containing 4 modules. Similarly, for $RB125$ network shown
in Fig.\ref{fig:subfig:2b}, $\Gamma(25)=0.71$ corresponding to 25
communities structure when $\tau=3$ is larger than $\Gamma(5)=0.18$
when $\tau=13$. The robustness of community structure indicated by
soft stability $\Theta$ favors finer but obvious modules which reasonable for many
real networks. In addition, the difference between the stability
measure we proposed and the modularity $Q$ \cite{Newman02} is emphasized. We also applied our framework to the hierarchical network with different modular sizes and some representative real networks. Finally, the relationships between our work and some famous concepts proposed in \cite{Delvenne} and \cite{Reichardt} are analyzed. These results are shown in the part of \textbf{Supplementary Material}\cite{Appendix}.

\section{6. Conclusion}
In summary, we have presented a more theoretically-based community detection framework
which is able to uncover the connection between network's community structures
and spectrum properties of Potts model's local uniform state.
Important information related to
community structures can be mined from a network's spectral
significance through a Markov process computation, such as the stability of modularity structures and the
optimal number of communities. Our method does not provide a unique optimal partition for the graph.
Rather, we obtain number of stable levels and stability at each level over different layers of the hierarchical structure.
Its effectiveness and efficiency have been demonstrated and verified both theoretically and experimentally.

\acknowledgments
The authors are separately supported by NSFC grants 11131009 and 71071090.

\end{document}